\documentclass[a4paper,12pt]{article}
\usepackage{epsfig}
\usepackage[dvips,usenames]{color}
\usepackage{graphicx}

\renewcommand{\textfraction}{0.0}
\renewcommand{\topfraction}{1.0}
\renewcommand{\bottomfraction}{1.0}   
\renewcommand{\baselinestretch}{1.5}

\newlength{\dinwidth}
\newlength{\dinmargin}
\newcommand{\Correct}[1]{{\color{Red}\fbox{\color{Black}#1}}}
\newcommand{\spur}[1]{\not\! #1 \,}
\begin{document}
\title{Semi-inclusive Decay $B\to \phi X_s$: Rate and Momentum Spectrum 
of $\phi$}
 
\bigskip

\author{
 Gad Eilam\footnote{eilam@physics.technion.ac.il}
\\
{ \small  \it Department of Physics, Technion-Israel Institute of
Technology,
 Haifa 32000,  Israel}\\
~and~ Ya-Dong Yang\footnote{yangyd@henannu.edu.cn}\\
{\small  \it Department of Physics, Henan Normal University, Henan
453002, P.R.China}
}
\maketitle

\bigskip\bigskip

\begin{abstract}
\noindent
We study the rate and 
$\phi$ momentum distribution in semi-inclusive decays $B \to\phi X_s $ 
induced by the quark level processes $b\to\phi s$ and $b\to\phi s g$,
in which 
the gluon is radiated from the internal charm quark loop
 or emitted from the virtual 
gluon of the strong penguin (inner bremsstrahlung).  We find 
${\cal B}(b\to\phi s )=6.7\times 10^{-5}$ and  ${\cal B}(b\to\phi s
g)=3.8\times 10^{-5}$.
The momentum spectrum of $\phi$ produced by $b\to\phi s g $ is very
broad. With the 
cut $|{\bf k}_{\phi}|\geq 2.0$ GeV,
${\cal B}(b\to\phi s )=6.1\times 10^{-5}$
(where the Fermi motion of the $b$-quark in the $B$-meson is described
by a Gaussian),
and ${\cal B}(b\to\phi s g)=1.0\times 10^{-5}$. Due to the
 special nature of
$\phi$, many difficulties which hindered a reliable 
theoretical prediction for 
$B\to \eta' X_s$ decay are absent in the process  $B\to \phi
X_s$.           
Therefore, theoretical predictions for $B\to \phi X_s$ are
relatively clean. 
Moreover, the clear experimental
signature of the $\phi$ is of great help.
Data for $B\to\phi X_s$, both the branching ratio and the $\phi$ 
momentum distribution, would
teach
us about the strength of strong penguins which might be of great
importance in
the search for CP violation and for new physics at $B$ factories.
\\
{\bf PACS Numbers 13.25.Hw, 12.38.Bx, 12.15Mm}
\end{abstract}

\newpage
\section{Introduction}
Understanding of 
pure penguin decays of $B$ mesons, is of utmost importance. Penguins
have been 
serving as powerful probes for the Standard
Model (SM) and 
for beyond the SM scenarios.
Measurements of the electromagnetic penguin process
$b\to s \gamma$ 
by CLEO~\cite{cleo1}, which agrees with the SM, has provided very
stringent
constraints on new physics. The 
strong penguin processes $B\to \eta' X_s$, $B\to \eta' K^{(*)}$, $B\to
\phi K^{(*)}$ 
have been observed by CLEO~\cite{cleo2}, BABAR~\cite{babar1,babar2}
and BELLE~\cite{belle1, belle2}. As is well known, 
it is quite difficult to provide theoretical estimates of 
exclusive nonleptonic decays, although there has been some
progress on this topic. In particular we refer to the 
QCD factorization~\cite{BBNS} and pQCD
approaches~\cite{lisanda} developed recently.
To explain the large yields of $\eta'$, we encounter  unknown parameters 
like the content of $\eta'$, mixing-angles, the $gg-\eta'$ coupling and so
on, which 
have hindered reliable theoretical predictions. There are also
suggestions that 
new physics could enhance the magnitude of the 
strong penguin. At present  
we cannot conclude how large 
the window is for new physics or whether it 
is required to explain the data.
The semi-inclusive $B\to \phi X_s$ decay is theoretically cleaner
than $B\to\eta' X_s$,
since $\phi$ is almost a pure $s\bar s$ state with mass larger
than $2M_{K}$
and 
does not couple to two gluons. In addition, 
$\phi$ coupling to three gluons is highly suppressed by the
OZI rule. 
 
There is a number of studies of $B\to K^{(*)}\phi $, using
different approaches
~\cite{eilamex, cheng2, he, lihn1}. Unfortunately the results do not
converge. 
QCD factorization predictions~\cite{cheng2,he} are smaller than 
those of pQCD~\cite{lihn1}.   Compared with exclusive decays, the
theoretical 
predictions for inclusive and semi-inclusive decay rates of $B$ mesons
rest on 
more solid grounds.  
The semi-inclusive decay $B\to\phi X_s$ was studied
a few years 
ago~\cite{eilam}. Recently it has caught renewed interest~\cite{he,
cheng} by generalizing 
the QCD factorization formalism~\cite{BBNS} to semi-inclusive processes. 
In the present paper, we will study both the branching ratio and the
$\phi$ momentum distribution, hereafter denoted by $|{\bf k}_\phi|$.
We take into
account the Fermi 
motion of the $b$-quark for $b\to s \phi$ and the $b\to s g^* g$
strong 
penguin effects arising from the 
inner bremsstrahlung processes in which a gluon is emitted from
the charm loop or from the virtual gluon in
$b\to s g^{*}$. 

Possible large $b\to s g g$ contribution to inclusive $B$ meson charmless
decays was 
discussed in Refs.~\cite{hou1, hou2, wyler, yao, greub}. Furthermore,
while studying penguin effects, 
Gerard and Hou~\cite{hou1, hou2}
found that the 
higher order processes $b\to s g g$ and $b\to s q\bar q$ dominate over
$b\to s g$.
Subsequently, 
it was found by Simma and Wyler~\cite{wyler}, and independently by Liu
and Yao~\cite{yao}, that 
$b\to s gg$ is considerably suppressed as compared to 
$b\to s q\bar q$.
Both groups  found that 
the 
large form factor
$F_{1}(x)$ is absent when both gluons are on-shell, while if
one of the two  
gluon goes off-shell, $F_1(x)$ survives.  
 
In the present paper we work within the 
framework of an effective low energy theory with five active quarks
which is 
obtained by integrating out heavy top and heavy gauge
bosons~\cite{buras}.  
We also use the QCD factorization framework  
to deal with the hadronic dynamics of $\phi$ formation. We find that
the $b\to \phi s g$
contribution is very significant,
 ${\cal B}(b\to \phi s g)=3.8\times 10^{-5}$.
After imposing a
momentum cut $|{\bf k}_{\phi}|\geq 2.0$ GeV,  ${\cal B}(b\to \phi s g)$ is
reduced to 
$1.0\times 10^{-5}$,
which is about $16\%$ of the fast $\phi$ production due to $b\to \phi s$. To
understand 
this "unusual" large contribution, we note that the amplitude for $b\to\phi
s g$ is 
characterized by  a factor $C_1 \frac{g_s^3}{16 \pi^2}\simeq 0.029 $ which
is numerically 
comparable to $C_6 =-0.041$, known to be the largest coefficient of the
strong penguin 
four Fermion operators. It is interesting to note that similar large
next-to-leading
(NLO) corrections have also been found for $b\to sg$~\cite{greub} and
$B\to K^{*}\gamma$
~\cite{buchalla}.      

The remainder of the paper is organized as follows: In Sec.2 we study the
momentum 
spectrum of $\phi$ resulting from $b\to\phi s$ by taking into account the
effect of Fermi motion. 
Sec.3 contains a calculation of the $b\to \phi s g$
contributions.
We present and discuss our results 
in Sec.4. Some useful
formulas 
are given in three Appendices.  
                
\section{ $B\to\phi X_s$ induced by $b\to\phi s$ }
In the low energy effective theory of the SM, the relevant 
effective Hamiltonian
is written as follows~\cite{buras}
\begin{equation}
{\cal H}_{eff}
=\frac{4G_{F}}{\sqrt{2}}
\left[ V_{cb} V_{cs}^* \sum_{i=1}^{2}
C_{i}O_{i}  -V_{tb} V_{ts}^* 
\left( C_g O_g +
\sum_{j=3}^{10}
C_{j}O_{i} 
\right)
\right].
\label{heff}
\end{equation}
The operators ${\cal O}$  are listed in App.A. 
The Wilson coefficients evaluated at scale $\mu=m_b$ are~\cite{buras}
\begin{equation}
\begin{array}{lll}
   C_1 =1.082, & C_2 =-0.185, & C_3 =  0.014, \\
  C_4= -0.035, & C_5=  0.009, & C_6= -0.041, \\
  C_7= -0.002/137,& C_8=  0.054/137, & C_9= -1.292/137,\\
  C_{10}= 0.262/137,& C_g =-0.143. & \\
\end{array}\label{ci}
\end{equation}
In the naive factorization approach, $b\to\phi s$ decay is a color-
suppressed 
process depicted in Fig.1. The amplitude can be easily written 
as
\begin{equation}
 {\cal M} = \frac{G_{F}}{\sqrt{2}} \bar s \gamma_{\mu} (1-\gamma_5 ) 
b\,\cdot\, \epsilon_{\phi}^{\mu} f_{\phi} m_{\phi} {\cal A}_{p},  
\end{equation}
where 
\begin{equation} 
{\cal A}_{p}= -V_{tb} V_{ts}^* 
\left( a_3 +a_4 +a_5 -\frac{1}{2}\left( a_7 +a_9 +a_{10} \right)
\right), 
\label{apnaive}
\end{equation}
and $a_{2n-1}=C_{2n-1}+C_{2n}/N_{c}$,
$a_{2n}=C_{2n}+C_{2n-1}/N_{c}$, $(n=2,3,4,5)$. 
\begin{figure} 
\begin{center}
\scalebox{1}{\epsfig{file=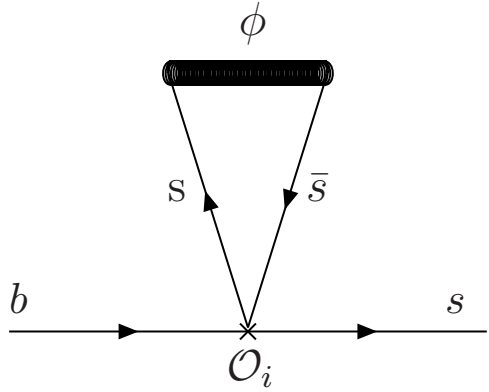}}
\caption{\small Leading diagram for  $(B\to \phi X_s)_2$ defined
in Eq.\ref{2body}. ${\cal
O}_{i} $ are strong
penguin operators in which top penguin effects are embedded.} 
\end{center}
\end{figure} 
The branching ratio is 
\begin{equation}
{\cal B}(b\to \phi s) =\tau_B \frac{G_F^2 f^2_{\phi} m_b^3 }{16 \pi} 
|{\cal A}_p |^2 \left( 1-\frac{m^2_{\phi}}{m^2_b } \right)^2 
 \left( 1+2\frac{m^2_{\phi}}{m^2_b } \right).  
\label{ap}
\end{equation}
Let us denote the two body contribution to $B\to\phi X_s$ by
\begin{equation}
{\cal B}\left[(B\to\phi X_s)_2\right] \equiv{\cal B}(b\to\phi s). 
\label{2body}
\end{equation}
Using $f_{\phi}=233$ MeV, $m_b =4.8$ GeV, 
and the Wolfenstein parameterization 
$V_{tb}=1$, $V_{ts}=-A \lambda^2$ with  $A=0.817$ and $\lambda=0.2237$
from 
Ref.~\cite{ckm2000}, we have 
\begin{equation}
{\cal B}\left[(B\to \phi X_s)_2\right]
=4.9 \times 10^{-5}~~~~~~~{\rm (naive~factorization).}
\end{equation}

\begin{figure}[htbp]
\begin{center}
\scalebox{0.9}{\epsfig{file=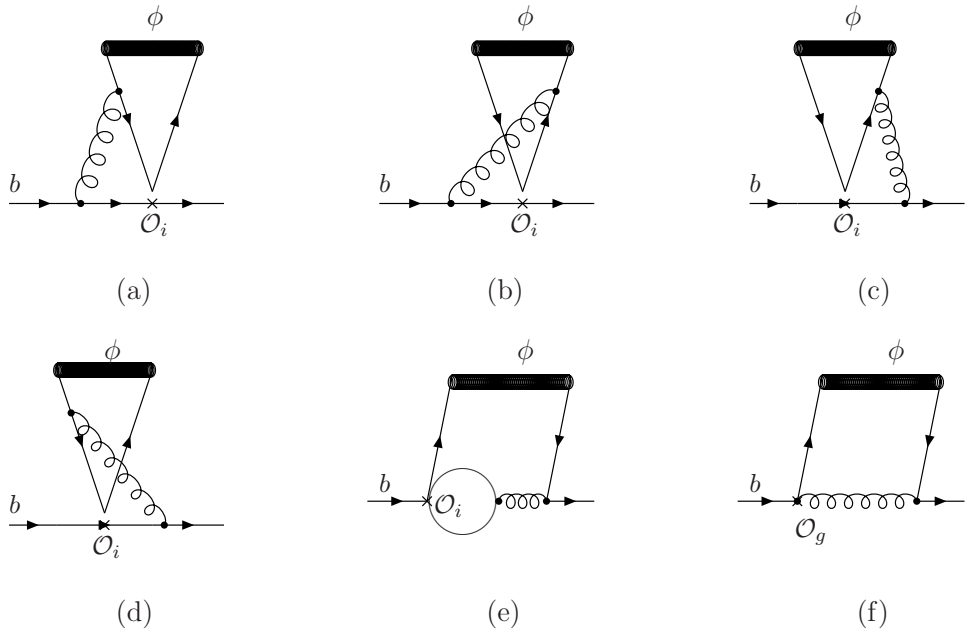}}
\caption{\small Subleading diagrams for  $(B\to \phi X_s)_2$ defined
in Eq.\ref{2body}.}  
\end{center}
\end{figure}
It is believed that naive factorization works very well for
color-allowed 
processes since Bjorken's color transparency argument~\cite{bj} applies,
while for 
color-suppressed processes nonfactorizable effects could play an important 
role. As demonstrated by Cheng and Soni~\cite{cheng}, the
nonfactorization effects 
depicted in Fig.2 could be calculated in the BBNS QCD factorization 
framework. The calculation of the effects of Fig.2 was carried out in 
Refs.~\cite{cheng,yangpv}. Incorporating  the nonfactorizable contributions,
${\cal A}_p$ in Eq.\ref{apnaive} is modified to 
\begin{eqnarray}
{\cal A}_p^{\prime}&=&V_{ub} V_{us}^{*} \left[ a_3^u +a_4^u +a_5^u 
-\frac{1}{2}\left( a_7^u +a_9^u +a_{10}^u \right)+a_{10a}^u \right]
\nonumber \\
&+& V_{cb} V_{cs}^{*} \left[ a_3^c +a_4^c +a_5^c 
-\frac{1}{2}\left( a_7^c +a_9^c +a_{10}^c \right)+a_{10a}^c \right], 
\end{eqnarray} 
where the coefficients $a_i^q$'s can be found in 
Ref.~\cite{yangpv}. We subtract the contribution of gluon-spectator
interactions ($f_{II}$ in~\cite{yangpv}),
choose $\gamma=54.8^{\circ}$ and present the numerical results in
Table 1. 
\ From Table 1, we can see that our results agree with those by Cheng and
Soni~\cite{cheng}.
\begin{table}[htbp]
\caption{\small Numerical values for $a_i^p$ (in units of $10^{-4}$) in QCD
factorization and 
in naive factorization. Our results are given for
$\gamma=54.8^{\circ}$. Cheng and Soni's results~\cite{cheng} are displayed
for comparison. 
}
\begin{center}
	\begin{tabular}{cccc}
		\hline  
            \hline
$a_i^q$   & Our results & Ref.~\cite{cheng} & Naive factorization  \\
\hline
$ a_3^{c,u}$ & 76.0+27.8i         &($a_3$ ) 74+26i  &   23         \\
$ a_4^{c}  $ &-375-71.6i          &($a_4$) -353-58i & -303\\
$a_4^{u}   $ &-318-151i           &                 & -303 \\
$a_5^{c,u} $ &-68.3-31.4i         &($a_5$) -67-30i  & -46\\
$a_7^{c,u} $ &1.37+0.3i          &($a_7$) -0.89-1.13i &1.16\\
$a_9^{c,u} $ &-90.8-1.4i          &($a_9$) -92.9-2.8i &-87.9\\
$a_{10}^{c,u}$ &1.95+7.23i         &($a_{10}$) 0.6+6.4i &12.2\\
$a_{10a}^{c,u}$&-0.52-1.0i         &                   &     \\
\hline
\hline
	\end{tabular}
\end{center}
\label{tab1}
\end{table}
We have 
\begin{equation}
{\cal B}\left[(B\to\phi X_s)_2\right]=6.7\times 10^{-5}~~~~~~~({\rm 
QCD~factorization}).
\label{2bodynocut}
\end{equation}
This large branching ratio enhances the feasibility 
of measuring the 
strength of the strong penguin to test the SM by studying this decay mode 
at BABAR and BELLE. 

To study the momentum spectrum, we employ the ACCMM model~\cite{accmm}
(for an earlier version see~\cite{ali1}).
In this model the bound state corrections to free $b$-quark decays are
incorporated
by attributing to the $b$-quark Fermi motion within the meson. The spectator
quark is 
handled as an on-shell particle with definite mass $m_{sp}$ and momentum
$|{\bf p}|=p$.
Consequently, the $b$-quark is considered to be off-shell with  a 
momentum dependent 
virtual mass $W(p)$ 
\begin{equation}
W^2 (p)=M^2_{B}+m_{sp}-2 M_{B} \sqrt{m^2_{sp}+p^2},
\label{fm1}
\end{equation}    
in which energy-momentum conservation is imposed. The momentum of 
the $b$-quark is
modeled by a 
Gaussian distribution function with a free parameter $p_F$
\begin{equation}
\phi_{F}(p)=\frac{4}{\sqrt{\pi}}\frac{1}{p^3_F}
\exp{\left(-\frac{p^2}{p^2_F} \right)}.  
\label{fm2}
\end{equation}  
It is interesting to note that the parameter $p_F$ and the $b$-quark average
mass 
$\langle m_b \rangle$ 
were obtained from a fit to the
photon momentum distribution in $B\to X_s \gamma$ by CLEO~\cite{cleopf},
to read $p_{F}=410$ MeV and $\langle m_b \rangle=4.690$ GeV. Using these values
and 
Eqs.\ref{fm1},~\ref{fm2},  we get $m_{sp}=298$ MeV for $B_{u,d}$.     

Now, the momentum spectrum of $\phi$ resulting from the decay 
$b\to\phi s$ of a $b$-quark of mass 
$W(p)$, is given by 
\begin{equation}
\frac{d\Gamma(|{\bf k}_{\phi}|,p) }{d|{\bf k}_{\phi}|}=
\frac{W(p)}{E_b} \frac{\Gamma_{0}(W(p))}{{\bf k}^b_{+}-|{\bf k}^{b}_{-}|}
\left[ 
\theta(|{\bf k}_{\phi}|- |{\bf k}^b_{-}|)-\theta(|{\bf k}_{\phi}|- 
{\bf k}^b_{+})\right],
\end{equation}   
where ${\bf k}^b_{+,-}$ provide the limits of the momentum range which results
from 
the Lorentz boost, $i.~e.$
\begin{equation}
{\bf k}^b_{\pm}(p)=\frac{1}{W(b)} (E_{b} k_0 \pm p E_{0}) 
\end{equation} 
with 
\begin{equation}
k_{0}=\frac{1}{2W(p)} (W(p)^2 -m^2_{\phi}), ~~
E_{0}=\sqrt{k^2_0 +m^2_{\phi}}, ~~ E_b =\sqrt{W(p)^2 +p^2}.
\end{equation}
Finally,
\begin{equation}
\Gamma_{0}(W(p))=\frac{G_F^2 f^2_{\phi} W(p)^3 }{16 \pi} 
|{\cal A}^{\prime}_p |^2 \left( 1-\frac{m^2_{\phi}}{W(p)^2} \right)^2 
 \left( 1+2\frac{m^2_{\phi}}{W(p)^2 } \right).  
\end{equation}
         
To get the momentum spectrum of $\phi$ from the semi-inclusive decay of
the $B$ 
meson, we have to fold in the $b$-quark momentum probability
distribution as 
given by $\phi_{F}(p)$, 
\begin{equation}
\frac{d{\cal B}(B\to\phi X_s)}{d|{\bf k}_{\phi}|}=\tau_{B} 
\int^{p_{max}}_{0} dp \phi_{F}(p) p^2  
\frac{d\Gamma(|{\bf k}_{\phi}|,p) }{d|{\bf k}_{\phi}|},
\end{equation}
where $p_{max}$ is determined by $W(p)^2 \geq (m_{\phi}+m_s)^2$, 
leading to
\begin{equation} 
p_{max}=\frac{1}{2M_{B}}\sqrt{\left(M^2_B +m^2_{sp}-(m_{\phi}+m_s )^2
\right)^2 
-4M^2_{B} m^2_{sp}}.
\end{equation}    
\begin{figure} 
\begin{center}
\scalebox{0.7}{\epsfig{file=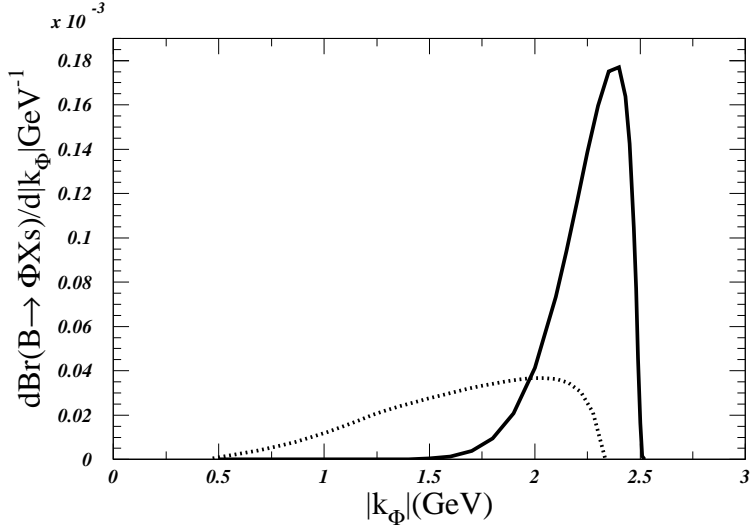}}
\caption{\small Momentum spectrum of $\phi$ in  $B\to \phi X_s$ decay.
The solid curve is for $b\to \phi s$ including Fermi
motion.
The dotted curve is for $b\to\phi s g$.}  
\end{center}
\end{figure}

Our numerical results for the $\phi$ momentum spectrum are presented in
Fig.3. 
We can see  the spectrum  peaking near $|{\bf k}_{\phi}|=2.4$ GeV which
corresponds 
to $M_{X_s}=1.15$ GeV.  To suppress indirect $\phi$ production, one can
impose a 
momentum cut $|{\bf k}_{\phi}|\geq 2.0$ GeV. With this cut,  the
branching ratio is 
\begin{equation}
{\cal B}\left[(B\to \phi X_s)_2\right]= 6.1\times 10^{-5}
~~~~~~~({\rm QCD~factorization+Fermi~motion+}
|{\bf k}_{\phi}|\geq 2.0~{\rm GeV}).
\label{2bodycut}
\end{equation}

\section{Contribution of the three body decay $b\to\phi s g$}
The $b\to \phi s g$ contribution to the momentum spectrum of $\phi$ 
has been studied by Deshpande $et~al$.~\cite{eilam} for the case of
gluon radiating from external quarks. They find that the effect is 
rather small, $ {\cal B}(b\to \phi s g)/{\cal B}(b\to \phi s )\approx
3\% $. From now on we will neglect gluon emission from external lines, 
and study instead the effect of  the gluon
radiating from an  
internal quark or splitting off an internal off-shell gluon as shown by Fig.4.
Both these processes are sometimes referred to as ``inner bremsstrahlung''. 
\begin{figure}[htbp]
\begin{center}
\scalebox{0.9}{\epsfig{file=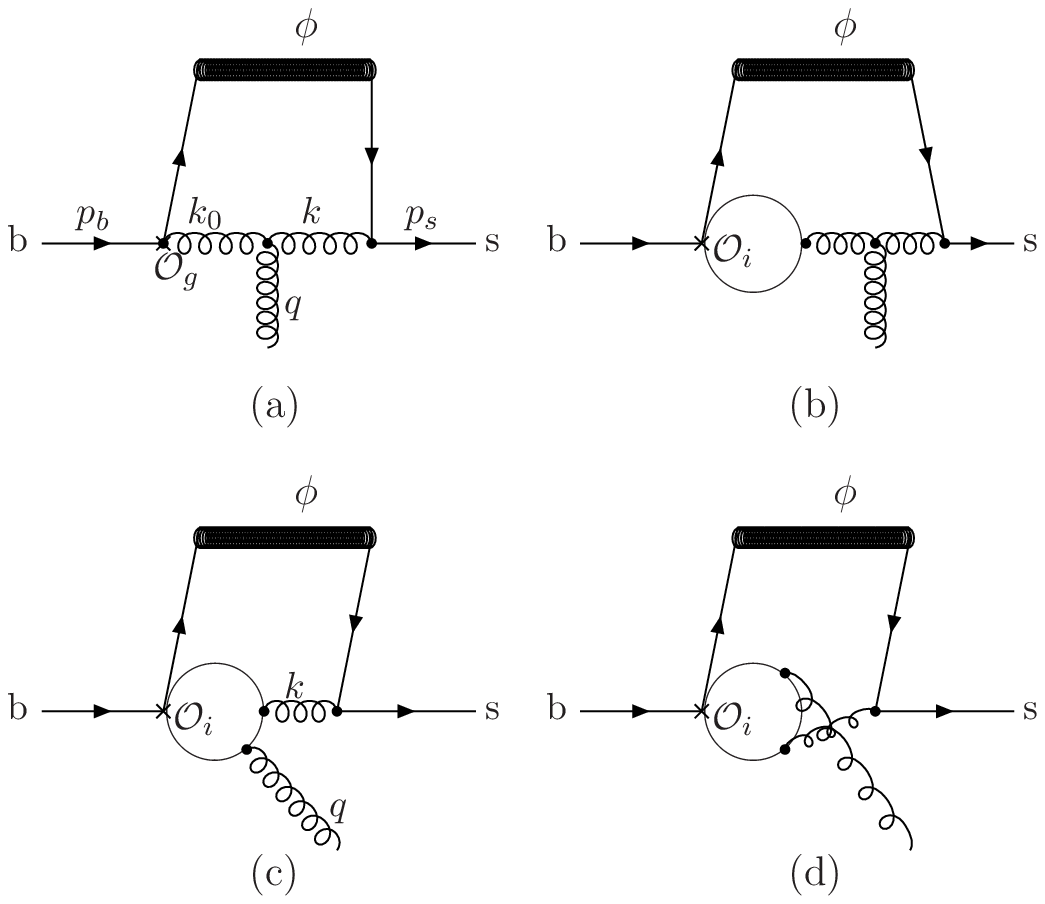}}
\caption{\small  Diagrams for 
 ${\cal B}\left[(B\to \phi X_s)_3\right]$ defined in Eq.\ref{3body}.}  
\end{center}
\end{figure} 
The calculation of Fig.4.(a)-(b) is straightforward. However the
calculation 
of Fig.4.(c) and (d) is tedious. A calculation of the charm 
loop for $b\to sg^{*}g$
has been
carried out recently by Greub nad Liniger~\cite{greub}, in a detailed
study of 
${\cal O}(\alpha_s)$ corrections to $\Gamma(b\to sg)$. It gives
\begin{equation}
J^{AB}_{\alpha\beta}=T^{+}_{\alpha\beta}(q,k)\{ T^{A}, T^{B}  \}
+T^{-}_{\alpha\beta}(q,k)[ T^{A}, T^{B}  ].
\end{equation}
$q,\,A$ and $\alpha$ denote the momentum, color and Lorentz index for the
on-shell 
gluon, respectively, while $k,\,B$ and $\beta$ stand for the same attributes
for the other, off-shell, gluon.  The
quantities 
$T^{+}_{\alpha\beta}(q,k)$ and $T^{-}_{\alpha\beta}(q,k)$ are
~\cite{greub}
\begin{eqnarray}
T^{+}_{\alpha\beta}(q,k)&=& \frac{g^2}{32\pi^2} \biggl[ 
E(\alpha,\beta, k)\triangle i_{5} + E(\alpha,\beta, q)\triangle i_{6} 
-E(\beta, k, q) \frac{k_{\alpha}}{q\cdot k}\triangle i_{23} 
\biggr. \nonumber \\
\biggl.
&&-E(\alpha, k, q) \frac{k_{\beta}}{q\cdot k}\triangle i_{25}
-E(\alpha, k, q) \frac{q_{\beta}}{q\cdot k}\triangle i_{26}
\biggr] L,\\
T^{-}_{\alpha\beta}(q,k)&=& \frac{g^2}{32\pi^2} \biggl[ 
\spur{k} g_{\alpha\beta}\triangle i_{2} +\spur{q}g_{\alpha\beta} 
\triangle i_{3}+\gamma_{\beta} k_{\alpha} \triangle i_{8} 
+\gamma_{\alpha} k_{\beta}\triangle i_{11}+\gamma_{\alpha} q_{\beta}
\triangle i_{12}     \biggr.  \nonumber \\
&& \biggl. 
+\spur{k}\frac{ k_{\alpha}k_{\beta} }{q\cdot k} \triangle i_{15} 
+\spur{k}\frac{ k_{\alpha}q_{\beta} }{q\cdot k} \triangle i_{17}
+\spur{q}\frac{ k_{\alpha}k_{\beta} }{q\cdot k} \triangle i_{19}
+\spur{q}\frac{ k_{\alpha}q_{\beta} }{q\cdot k} \triangle i_{21}
\biggr]L.    
\end{eqnarray}
The dimensionally regularized expressions for the $\Delta i$ functions 
are given by Greub and Liniger~\cite{greub}.   
To reduce the difficulty in numerical calculations of
multiple integrals,
we specialize their functions  to $d=4$ and integrate out 
the two Feynman parameters. For $d=4$, $E(\alpha,\beta,\gamma)=-i
\epsilon_{\alpha\beta\gamma\mu}\gamma^\mu \gamma_5$, in the
Bjorken-Drell conventions. 
The analytical expressions for the
functions $\triangle i$ are 
collected in App.B, where we have used the $\overline{{\rm MS}}$ 
scheme. 

The dimensionless variables $s,\,t$ and $u$ are defined as
\begin{equation}
s=\frac{(p_{\phi}+p_s )^2}{m_b^2},~~~
t=\frac{(p_{\phi}+q )^2}{m_b^2},~~~
u=\frac{(q+p_s )^2}{m_b^2}.
\end{equation}
We write the amplitude of the three body contribution to $B\to\phi X_s$ as 
\begin{equation}
{\cal B}\left[(B\to\phi X_s)_3\right] \equiv{\cal B}(b\to\phi sg). 
\label{3body}
\end{equation}
Now, neglecting gluon bremsstrahlung from external quarks, the inner 
bremsstrahlung diagrams (see Fig.4), lead to the matrix element 
\begin{equation}
{\cal M}_3 ={\cal M}_{O_g}+{\cal M}_{O_c} +{\cal M}_{\triangle_+}
 +{\cal M}_{\triangle_-}, 
\end{equation}
and its square reads 
\begin{eqnarray}
\mid{\cal M}_3 \mid^2 &=&| {\cal M}_{O_g} |^2 +| {\cal M}_{O_c} |^2 +
|{\cal M}_{\triangle_+}|^2  +|{\cal M}_{\triangle_-}|^2 
+2 \Re( {\cal M}_{O_g} {\cal M}_{O_c}^{\dagger} )
+2 \Re({\cal M}_{\triangle_+}{\cal M}_{\triangle_-}^{\dagger} )
\nonumber \\
&&+2 \Re ({\cal M}_{\triangle_+}{\cal M}_{O_g}^{\dagger} )
+2 \Re ({\cal M}_{\triangle_+}{\cal M}_{O_c}^{\dagger} )
+2 \Re ({\cal M}_{\triangle_-}{\cal M}_{O_g}^{\dagger} )
+2 \Re ({\cal M}_{\triangle_-}{\cal M}_{O_c}^{\dagger} ).
\label{m3}
 \end{eqnarray}
The explicit forms for the terms in the amplitude squared are given in 
App.C. 

The decay distribution of the $\phi$ in 
${\cal B}\left[(B\to \phi X_s)_3\right]$
is
\begin{equation}
\frac{d \Gamma(b\to\phi s g)}{du\,dt} =\frac{1}{256 \pi^3} m_b
\frac{G^2_F}{2}
|V_{ts}|^2 \frac{1}{6} |{\cal M}_3|^{2},
\end{equation}
where the factor $\frac{1}{6}$ is
due to the average of spin and color of the $b$-quark.
In the phase space integration, we impose the cut $(p_s +q)^2 \geq m_K^2$. 

\section{Results and discussion}
Our results for the contribution of $b\to\phi s g$ to the $\phi$ spectrum, 
are displayed in Fig.3 by the dotted 
curve, which is much broader than that due to $b\to
\phi s$. One can improve the predictions for the spectrum of the $\phi$
mesons by adding Fermi motion to the three body processes, including
radiation from external quarks.
The contribution 
of $b\to\phi s g$ is quite substantial, 
\begin{equation}
{\cal B}\left[(B\to \phi X_s)_3\right] =3.8\times10^{-5}, 
\label{3bodynocut}
\end{equation}
where the left hand side of the last equation was defined in Eq.\ref{3body}.
Now, after applying a momentum
cut
$|{\bf k}_{\phi}|\geq 2.0$ GeV, the branching ratio is reduced to 
\begin{equation}
{\cal B}\left[(B\to \phi X_s)_3\right] =1.0\times10^{-5}~~~~~~~ 
(|{\bf k}_{\phi}|\geq 2.0~{\rm GeV}).
\label{3bodycut}
\end{equation}
which is about $16\%$ of
${\cal B}(b\to\phi s)$.
Interestingly, such a large NLO correction was also found in $b\to
s g$~\cite{greub} and 
$B\to K^{*} \gamma$~\cite{buchalla}. In Ref.~\cite{greub}, Greub and
Liniger have found  
that the next-to-leading lorgarithmic result  
${\cal B}^{NLL}(b\to sg)=(5.0\pm1.0)\times 10^{-5}$ is more than a
factor of 
two larger than the leading logarithmic result
${\cal B}^{LL}(b\to sg)=(2.2\pm 0.8)\times 10^{-5}$. The reasons for
this 
enhancement also apply here.

It is well known that the strength of strong penguins is an important
issue of current interest, relevant to CP violation,
tests of the SM and to searches for new physics signals at BABAR and BELLE. The
first evidence
for a strong penguin was found  by CLEO in 1997 by measuring $B\to K \eta'$
and 
$B\to \eta' X_s$~\cite{cleo97},  which turned out be "unexpectedly large". 
Recently both BABAR~\cite{babar1} and BELLE~\cite{belle1}
have confirmed the CLEO measurement with improved precision.
Theoretically, 
there are many models to explain the yield of $\eta'$ 
in $B$ decays. The measurement of the $\eta'$ momentum spectrum by 
CLEO~\cite{babarspec} has ruled out models with a $b\to c{\bar c}s$ 
enhancement through a possible $c\bar c$ content in the $\eta'$ wave
function.
Up to now, there are two surviving explanations
for the high yield of $\eta'$ in $B$ decays: A model in which new 
physics enhances the strong penguin $b\to s g$~\cite{kagan} and a model
incorporating the QCD 
anomaly coupling $g-g-\eta'$~\cite{ggeta, yang}. The $\eta'$ is a
very complicated           
object in QCD. There are unsolved puzzles involving the $\eta'$, such as its
large mass,
mixing between flavor singlet and octet, gluonic  content, decay
constants and its mixing 
with glueballs, which inhibit reliable theoretical predictions
for $B\to\eta' K^{(*)}$
and $B\to\eta' X_s$. 
The aforementioned difficulties are
absent for $\phi$ and  
theoretical predictions for  $B\to \phi X_s$ are rather clean.
Furthermore,
$\phi$ has a clear experimental signature. We can therefore conclude
that  the theoretical study and
experimental 
measurement  of $B\to \phi X_s$  will provide quite a clean ground for
testing the SM and 
searching for new physics.
 
In summary, we have studied the semi-inclusive decays $B\to \phi X_s$.
We have 
calculated  the contributions of $b\to \phi s$ and of the 
$b\to \phi s g$ subprocesses with the
gluon
radiated from the charm loop or from the
off-shell 
gluon of $b\to s g^{*}$. These effects are found to be quite large, giving 
${\cal B}(b\to \phi s g)=3.8\times 10^{-5}$. Cutting the $\phi$ momentum
$|{\bf k}_{\phi}|\geq 2.0$ GeV, results in 
${\cal B}(b\to \phi s g)=1.0\times 10^{-5}$.
Adding this contribution to the dominant process $b\to\phi s$, we
predict 
\begin{equation} 
{\cal B}(B\to \phi X_s)=10.5\times 10^{-5}
\end{equation}
and
\begin{equation} 
{\cal B}(B\to \phi X_s)=7.1\times 10^{-5}~~~~~~~(|{\bf k}_{\phi}|\geq
2.0~{\rm GeV}).       
\end{equation}
This large decay rate and its clear signature render detailed 
studies of $B \to \phi X_s$ 
at BABAR and BELLE feasible in the near future. These will 
shed light on the strength of strong penguins. 

\section*{Acknowledgments}

This work is supported by the US-Israel Binational Science
Foundation, the Israel Science Foundation and the
Fund for Promotion of Research at the Technion. 
YDY would like to thank the members of the HEP group at Technion for 
their support and hospitality during his stay there. 
\section*{Appendix A: The operators $O_{i}$}
The operators in Eq.\ref{heff} read 
\begin{equation}\begin{array}{llllll}
O_1 & = & \bar c_\alpha\gamma^\mu L b_\alpha\cdot 
\bar s_\beta\gamma_\mu L c_\beta\ ,   &
O_2 & = & \bar c_\alpha\gamma^\mu L b_\beta\cdot \bar 
s_\beta\gamma_\mu L c_\alpha\ , \\
O_3 & = & \bar s_\alpha\gamma^\mu L b_\alpha\cdot \bar
 s_\beta\gamma_\mu L s_\beta\ ,   &
O_4 & = & \bar s_\alpha\gamma^\mu L b_\beta\cdot \bar 
s_\beta\gamma_\mu L s_\alpha\ , \\
O_5 & = & \bar s_\alpha\gamma^\mu L b_\alpha\cdot \bar 
s_\beta\gamma_\mu R s_\beta\ ,   &
O_6 & = & \bar s_\alpha\gamma^\mu L b_\beta\cdot \bar 
s_\beta\gamma_\mu R s_\alpha\ , \\
O_7 & = & \frac{3}{2}\bar s_\alpha\gamma^\mu L b_\alpha\cdot 
e_{s}\bar s_\beta\gamma_\mu R s_\beta\ ,   &
O_8 & = & \frac{3}{2}\bar s_\alpha\gamma^\mu L b_\beta\cdot 
e_{s}\bar s_\beta\gamma_\mu R s_\alpha\ , \\
O_9 & = & \frac{3}{2}\bar s_\alpha\gamma^\mu L b_\alpha\cdot 
e_{s}\bar s_\beta\gamma_\mu L s_\beta\ ,   &
O_{10} & = & \frac{3}{2}\bar s_\alpha\gamma^\mu L b_\beta\cdot 
e_{s}\bar s_\beta\gamma_\mu L s_\alpha,\\
O_{g} & = & \frac{g_s}{16 \pi^2} \bar s_{\alpha}\sigma^{\mu\nu}R  
T^{a}_{\alpha\beta} m_b b_{\beta} G^{a}_{\mu\nu}.  & & &  
\label{operators}
\end{array}
\end{equation}
Where $\alpha$ and $\beta$ are the $SU(3)$ color indices and 
$L=(1 - \gamma_5)/2$, $R= (1 + \gamma_5)/2$.

\section*{Appendix B: Functions for the charm loop}
Using the notations $r_{1}=\frac{2q\cdot
k}{m_c^2},~~r_{2}=\frac{k^2}{m_c^2}$,  
$q$ for the momentum of the on-shell gluon in the final state and $k$ for the
momentum of
the off-shell gluon,  we have  
\begin{eqnarray}
\triangle i_{5}&=&-2+\frac{16+2r_2}{r_1}(G_{0}(r_2)-G_{0}(r_1 +r_2 ))
-\frac{12}{r_1}\left( G_{-}(r_2 )-G_{-}(r_1 +r_2 ) \right), \\
\triangle i_{6}&=&\frac{7}{4}+\frac{7}{r_{1}}+\frac{r_2}{2 r_1}
- \frac{r_2^2}{3r_1^2}
+\frac{r_2^2}{3r_1}-\frac{4r_2 (1+r_2 -r_1 r_2 )}{r_1^2}G_{0}(r_2
)\nonumber \\
&&+\frac{2(r_1 +3r_1^2 +r_2 +r_1 r_2)}{r_1^2 }G_{0}(r_1 +r_2 )
  -\frac{4(r_1 -3r_2 )}{r_1^2}G_{-}(r_2 )+
\frac{4(1 -3r_2 )}{r_1 }G_{-}(r_1 +r_2 ) \nonumber \\
&&+\frac{5(4-r_2 )r_2 }{r_1^2 }T_{0}(r_2 )
+\frac{5(r_1^2 +2r_1 (r_2 -2)+r_2 (r_2 -4))}{r_1^2} T_{0}(r_1 +r_2 ), \\
\triangle i_{23}&=&-2+\frac{4}{r_1}(G_{-}(r_2 )-G_{-}(r_1 +r_2 ))
  -\frac{2 r_2}{r_1}(G_{0}(r_2 )-G_{0}(r_1 +r_2 )), \\
\triangle i_{25}&=&-2(G_{0}(r_2 )-G_{0}(r_1 +r_2 ) ),\\ 
\triangle i_{2}&=&-\frac{8}{3}\ln\frac{\mu}{m_c}+
              \frac{22}{9}+ \frac{16+2 r_2}{3r_1}G_{0}(r_2 )
             -\frac{16-4r_1 +2 r_2}{3r_1}G_{0}(r_1 +r_2 ) \nonumber\\
           && -\frac{4}{r_1}(G_{-}(r_2 )-G_{-}(r_1 +r_2 )), \\
\triangle i_{3}&=&4\ln\frac{\mu}{m_c}-\frac{85}{36}+\frac{19-3r_2}{r_1}
               +\frac{r_2^2}{9r_1}\left( \frac{1}{r_1}-1\right)
               +\frac{4r_2}{3r_1}\left(1-r_2
-\frac{3}{r_1}\right)G_{0}(r_2 )
          \nonumber\\
           &&-\left(
                   \frac{10}{3}+\frac{2}{3r_1}-\frac{14r_2}{3r_1^2}
               +\frac{2r_2}{r_1}-\frac{4r_2^2}{3r_1^2} 
                                            \right)G_{0}(r_1 +r_2 )
           +\frac{4}{r_1}\left( 1+\frac{2r_2}{r_1} \right)G_{-}(r_2) 
       \nonumber \\
        &&-\frac{4}{r_1}\left( 1+\frac{3r_2}{r_1}-r_2 \right)G_{-}(r_1
+r_2 )
           +\frac{9r_2}{r_1^2}( 4-r_2 )T_{0}(r_2 )
       \nonumber \\
        &&+9\left( 1-\frac{4}{r_1}-\frac{4r_2}{r_1^2}+\frac{2r_2}{r_1} 
                 +\frac{r_2^2}{r_1^2} \right)T_{0}(r_1 +r_2 ), \\
\triangle i_{8}&=&\frac{16}{3}\ln\frac{\mu}{m_c}
               -\frac{32}{9}+\frac{8}{3r_1}(r_2 -4)G_{0}(r_2 )
               -\frac{8}{3}\left(
1-\frac{4}{r_1}+\frac{r_2}{r_1}\right)G_{0}(r_1 +r_2 ),\\
\triangle i_{11}&=&-\frac{8}{3}\ln\frac{\mu}{m_c}+\frac{22}{9}
               +\frac{4}{3}G_{0}(r_2 )+\frac{2(8+r_1 +r_2 )}{3r_1}
            \left( G_{0}(r_2 )-G_{0}(r_1 +r_2 )\right)\nonumber \\
         &&-\frac{4}{r_1}\left( 2G_{-}(r_1 )-G_{-}(r_2 )-G_{-}(r_1 +r_2
)\right),\\
\triangle i_{12}&=&-\frac{16}{3}\ln\frac{\mu}{m_c }+4+\frac{24-2r_2
}{r_1 }
              -\frac{2}{r_1^2}\left[ (1+2r_1 )r_2 G_{0}(r_2 )-(r_1^2
+r_2 +2r_1 r_2 )
              G_{0}(r_1 +r_2 )\right] 
              \nonumber \\
     &&   +\frac{8r_2 }{r_1^2 }(G_{-}(r_2 )-G_{-}(r_1 +r_2 ) )
\nonumber\\
     &&  +\frac{12}{r_1^2 }\left[ r_2 (r_2 -4)T_{0}(r_2 )+(r_1^2 +2r_1
(r_2 -2) 
       +r_2 (r_2 -4))T_{0}(r_1 +r_2 )\right],\\
\triangle i_{15}&=&-\frac{4}{3}\left(1+\frac{2}{r_2}\right)G_{0}(r_2)
                   +\frac{4}{3}\left(1+\frac{2}{r_1
+r_2}\right)G_{0}(r_1 +r_2 ),\\
\triangle i_{17}&=& -\frac{2}{3}-\frac{2(8+r_2 )}{3r_1} G_{0}(r_2 )
               +\frac{2}{3}(\frac{8+r_2}{r_1}+\frac{4}{r_1 +r_2})
               G_{0}(r_1 +r_2 ) \nonumber\\
       &&+\frac{4}{r_1}(G_{-}(r_2 )
        -G_{-}(r_1 +r_2 ) ), \\
\triangle
i_{19}&=&-\frac{2}{3}+\frac{2}{3}(3+\frac{r_2}{r_1}-\frac{10}{r_1} )
                    (G_{0}(r_2 )-G_{0}(r_1 +r_2 ))
               +\frac{4}{3(r_1 +r_2 )}
               G_{0}(r_1 +r_2 ) \nonumber\\
       &&+\frac{4}{r_1}(G_{-}(r_2 )
        -G_{-}(r_1 +r_2 ) ), \\
\triangle i_{21}&=&\frac{2}{3}+\frac{16}{r_1}+\frac{8}{3}\frac{r_2}{r_1}
       -\frac{2 r_2 (24r_1 +3r_1^2 +20r_2 +7r_1 r_2 +4 r_2^2 )}{3 r_1^2
(r_1 +r_2 )}
   G_{0}(r_1 +r_2 )
    \nonumber \\
&&+\frac{2 r_2}{3r_1^2} (20+3r_1 +4 r_2 )G_{0}(r_2 )+
\frac{4}{r_1^2 }\left[ 
             (r_1 +4r_2 )(G_{-}(r_1 +r_2 )-G_{-}(r_2 ))  
                \right.  \nonumber \\
   &&  \left. +2r_2 (4-r_2 )T_{0}(r_2 )+2(r_1 +r_2 )(r_1 +r_2
-4)T_{0}(r_1 +r_2 )
         \right].
\end{eqnarray}

The function $G_i (t) (i=-1,0,1,2)$ are defined as
\begin{equation}
G_{i}(t)=\int^1_0 dx x^i \ln[1-tx(1-x)-i\delta].   
\end{equation}
$G_{1,2}(t)$ can be simplified to $G_{0}$. The explicit form of
$G_{-1,0}(t)$ 
could be found in Ref.~\cite{greub}. In the equations above we
used $G_-(t)\equiv G_{-1}(t)$.

$T_{0,1}$ are defined by
\begin{equation}
T_{i}=\int^1_0 dx \frac{x^i}{1-x(1-x)t-i\epsilon} 
\end{equation}
Their explicit forms are given by 
\begin{eqnarray}
T_{0}(t) &=&\left\{ 
\begin{array}{cc}
\frac{ 4\arctan\sqrt{ \frac{t}{4-t}}}{\sqrt{t(4-t)}} ; &0\leq t \leq 4
\\
\frac{2i\pi+2\ln( \sqrt{t}-\sqrt{t-4} )-2\ln( \sqrt{t}+\sqrt{t-4})}
{\sqrt{t(t-4)}};  &t > 4.
\end{array} \right. 
\end{eqnarray}

The one loop two-point charm penguin function is well known  in the
form of 
\begin{equation} 
G(t)=-\frac{2}{3}+\frac{4}{3} \ln\frac{\mu}{m_c}-\int^1_0 d\xi
4\,\xi(1-\xi)
\ln[1-\xi(1-\xi)t+i\delta],   
\end{equation} 
whose explicit form is given by
\begin{equation}
G(t)=-\frac{8}{9}+\frac{4}{3} \ln\frac{\mu}{m_c}
-\frac{2}{3}\left(1+\frac{2}{t}\right) G_0 (t). 
\end{equation}

\section*{Appendix C: Squares and interferences of amplitudes}
Now we write down the terms in Eq.\ref{m3} needed for calculating the
inclusive decays 
\begin{eqnarray}
| {\cal M}_{O_g} |^2 &=& \left( \frac{g_s^3}{8\pi^2}\right)^2 
                       \left( \frac{f_{\phi}}{4} \right)^2
                   |C_8|^2 8 m_b^2 
           \biggl[ 4|F_1 |^2 s t^2 (s+t)+2\Re(F_0 F_1^*)tu(-7s^2
+6st+5t^2 )
           \biggr. \nonumber \\
    &&  \biggl. 
             +|F_0|^2 u \Bigl( 12st(t+u)+4s^2 (7t+u)+t^2 (t+11u) \Bigr)
      \biggr]. \\
| {\cal M}_{O_c} |^2 &=& \left( \frac{g_s^3}{8\pi^2}\right)^2 
                       \left( \frac{f_{\phi}}{4} \right)^2
                   |C_1|^2 8 m_b^2 
           \biggl[ |F_3 |^2 \Bigl( s^3+4t^2 u+s^2 (u-2t)+st(t+5u) \Bigr)
               \biggr. \nonumber \\
    &&  \biggl. 
         -2\Re(F_3 F_5^*) s(s-t)+ |F_5|^2 s (s+t)
       \biggr].
\end{eqnarray} 

For the charm loop penguin contributions, we get 
\begin{eqnarray}
| {\cal M}_{\triangle_+} |^2 &=& \frac{49}{81}\left(
\frac{g_s^3}{16\pi^2}\right)^2 
				\left( \frac{f_{\phi}}{4} \right)^2
                   		|C_1|^2 2 m_b^2 
                  \biggl\{ 
              4 s^2 (s+u)|J^0_{i5}|^2 +4t \Bigl( tu+s(t+2u) \Bigr)
|J^0_{i6}|^{2}
        \biggr. \nonumber \\
         &&+tu^2 \Bigl(2st+u(s+t)\Bigr) |J^0_{i23}|^2+t(2s^{3} -st^2
-t^3 )|J^1_{i23}|^2 
         \nonumber \\ 
       &&+\Re \biggl[ 4J^{0}_{i5} \Bigl( 2s^2 (s+t) J^{1*}_{i5}
         +2s(tu+st+su)J^{0*}_{i6}+stu(s+u)J^{0*}_{i23}+s^3 u
J^{1*}_{i23}
         \Bigr.  
                                              \nonumber \\
   && \Bigl.  -s^2 t(s+t)J^{2*}_{i23}  \Bigr)+4sJ^{1}_{i5} 
       \Bigl( 2t(s+t)J^{0*}_{i6} -(s^2 -st-t^2 ) u J^{0*}_{i23}     
              +st (s+t) J^{1*}_{i23} \Bigr)   \nonumber \\ 
   && +4J^{0}_{i6} \Bigl( u \bigl( t^2 u-s(t+u)(s-t) \bigr) J^{0*}_{i23}
     +2 s^2 tu J^{1*}_{i23}-st^2 (s+t) J^{2*}_{i23} \Bigr)          
                                                \nonumber\\
     &&\biggl. \biggl. 
     +2 su J^{0}_{i23} \Bigl( 
   s(tu-st-su) J^{1*}_{i23}+t(s^2 -st-t^2 )J^{2*}_{i23}\Bigr)
    -2 s^2 t^2 (s+t)J^{1}_{i23} J^{2*}_{i23} \biggr]
           \biggr\} \\  
| {\cal M}_{\triangle_-} |^2 &=& \left( \frac{g_s^3}{16\pi^2}\right)^2 
				\left( \frac{f_{\phi}}{4} \right)^2
                   		|C_1|^2  4m_b^2 \nonumber \\
                 &\times & \biggl\{
            2s^2 (s+u)|J^0_{i2}|^2  +2st(t+u)|J^0_{i3}|^2
+2tu(s+t)|J^0_{i12}|^2
            +4st\Re\left(s J^0_{i2}J^{0*}_{i3}-u  J^{0}_{i8}J^{0*}_{i12}
\right)
           \nonumber \\
         &&-4s^2 (s+t)\Re J^{0}_{i8}J^{1*}_{i8} 
          -2s^2 tu \Re\Bigl[ J^0_{i8}J^{1*}_{i17}-J^0_{i8}J^{1*}_{i21}+
          J^1_{i8}J^{0*}_{i17}-J^1_{i8}J^{0*}_{i21}
\Bigr]                    
         \nonumber \\
     &&-stu\Re\Bigl[ 2(s+u) J^{0}_{i12}J^{0*}_{i17} +2t
J^{0}_{i12}J^{0*}_{i21}
                 +s(s+u) J^{1}_{i17}J^{0*}_{i17}\Bigr]
        \nonumber\\
     &&\biggr. -s^2 t^2 u\Re \Bigl[
J^{1}_{i17}J^{0*}_{i21}+J^{0}_{i17}J^{1*}_{i21}
          \Bigr] 
        \biggr\}.                          
\end{eqnarray}

The interference terms are read as
\begin{eqnarray}
 \Re( {\cal M}_{O_c} {\cal M}_{O_g}^{\dagger} )&=&
        -\left(\frac{g_s^3}{8\pi^2}\right)^2 |C_1 C_8|^2 
         \left(\frac{f_{\phi}}{4}\right)^2 8m_b^2 
        \Re \biggl\{ 
         2st [F_{3}(s-t)-F_{5}(s+t)]F_{1}^* ] 
            \biggr. \nonumber \\
 && \biggl.     
  -F_{0}^* [8 F_4 s^2 (s+t)+F_3 u (2s^2 -9st-3t^2 )+F_5 u (3st+t^2 -2s^2
)]
        \biggr\}, \\
\Re( {\cal M}_{\Delta_+} {\cal M}_{\Delta_-}^{\dagger} )&=&
        \frac{7}{9}\left(\frac{g_s^3}{16\pi^2}\right)^2 C_1^2 
         \left(\frac{f_{\phi}}{4}\right)^2 2 m_b^2  \nonumber \\ 
        &&\Re \biggl\{ 2s J^0_{i5} \Bigl[ 
                   2s^2 (J^{0*}_{i2}+J^{0*}_{i8}
)+2st(J^{0*}_{i3}+J^{1*}_{i8})
                +2u(s+t) J^{0*}_{i2}+stu (J^{1*}_{i17}-J^{1*}_{i21})
                                 \Bigr]
            \biggr. \nonumber \\
        &&-2s^2 J^{1}_{i5} \Bigl[ 
             2(s+t) J^{0*}_{i8}+tu (J^{0*}_{i17} -J^{0*}_{i21} ) \Bigr]  
        +4 J^{0}_{i6} \Bigl[
           st^2 J^{0*}_{i3}+(st(s+t)-s^2 u) J^{0*}_{i8}
                      \Bigr. \nonumber \\    
         &&\Bigl.  +t^2 u J^{0*}_{i12} 
           +st(s+u) J^{0*}_{i2} \Bigr]+2stu J^{0}_{i6} \Bigl[ 
           s
J^{0*}_{i17}+(t+u)J^{0*}_{i21}+t(J^{1*}_{i17}-J^{1*}_{i21})\Bigr]
       \nonumber\\
     && +u J^{0}_{i26} \Bigl[ 
         2s(s^2 -st-t^2 )J^{1*}_{i8}-2t(s^2 -su+tu) J^{0*}_{i12}
         -st(s^2 +tu)J^{1*}_{i17}\Bigr. 
     \nonumber \\
     &&\Bigl. 
     -2st^2 J^{0*}_{i3} -st(st+su-ut) J^{0*}_{i21}-2st(s+u) J^{0*}_{i2}
      \Bigr] -J^{1}_{i26}\Big[2s^2 t(s+t) J^{1*}_{i8} \Bigr. 
      \nonumber \\
     &&-2su(s^2+st+t^2 ) J^{0*}_{i8}   +2t^2 u(2s+t)J^{0*}_{i12}
      +stu(s^2 -tu)J^{0*}_{i17} \nonumber \\
     &&\biggl. \Bigl. 
       +s^2 t^2 u (J^{1*}_{i17}-J^{1*}_{i21})+
      stu (tu+st+su)J^{0*}_{i21} \Bigr]
     \biggr\}, \\                  
\Re( {\cal M}_{\Delta_+} {\cal M}_{O_c}^{\dagger} )&=&
        \frac{7}{9}\frac{g_s^6}{128\pi^4} C_1^2 
         \left(\frac{f_{\phi}}{4}\right)^2 4 m_b^2   
         \Re \biggl\{-F_{3}^{*} \Bigl[(s^2 +2tu+su-st)( s J^{0}_{i5}+t
J^{0}_{i6} )
         \Bigr. \biggr. 
      \nonumber \\
   && \Bigl. 
     +2stu( s J^{0}_{i23}+t J^{1}_{i23} )+ J^{0}_{i26} \Bigl( s(t+u)-s^2
-2tu \Bigr) 
     -2t^2 u(s+t) J^{1}_{i26} 
                \Bigr] \nonumber \\
    && 
      +2F_{4}^{*} s \Bigl[ (s+t)(2s J^{0}_{i5} +2t J^{0}_{i6} +st
J^{1}_{i23})
        -u\Bigl( s^2 J^{0}_{i23} +(st+t^2 ) J^{0}_{i26} \Bigr)
\Bigr]     
      \nonumber \\
    &&\biggl.  \Bigl.
      +F_{5}^{*} s(2sJ^{0}_{i5}+2tJ^{0}_{i6} -tu J^{0}_{i26}  ) 
      \Bigr] \biggr\}, \\
\Re( {\cal M}_{\Delta_+} {\cal M}_{O_g}^{\dagger} )&=&
        \frac{7}{9}\frac{g_s^6}{128\pi^4} |C_1 C_8 |
         \left(\frac{f_{\phi}}{4}\right)^2 4 m_b^2   
         \Re \biggl\{F_{1}^{*} st\Bigl( 4 s J^{0}_{i5}+4t J^{0}_{i6} 
           +3su J^{0}_{i23}-2tu J^{0}_{i26}   \Bigr)
        \biggr. 
      \nonumber \\
   &&-F_{0}^{*} \Bigl[ 4s^2 (s+t) (t J^{2}_{i23}-2J^{1}_{i5}  ) 
     -s^2 u(4s +t) J^{1}_{i23} +4us(s-t) J^{0}_{i5}        
               \Bigr. \nonumber \\
   && +(4s^2 +6st -t^2)tu J^{1}_{i26} -2(4s^2 -st+t^2 )u J^{0}_{i6} 
       +stu^2 J^{0}_{i23} \nonumber \\
&& \biggl.  \Bigl.
       +u^2 t^2 (t-4s ) J^{0}_{i26} \Bigr]     
      \biggr\}, \\
\Re( {\cal M}_{\Delta_-} {\cal M}_{O_g}^{\dagger} )&=&
        -\frac{g_s^6}{128\pi^4}  |C_1 C_8 | 
         \left(\frac{f_{\phi}}{4}\right)^2 4 m_b^2   
         \Re \biggl\{F_{0}^{*} \Bigl[2su(2s-t) J^{0}_{i2}-2tu (2s+t)
J^{0}_{i3}
           +2stu  J^{0}_{i8} 
         \Bigr. \biggr. 
      \nonumber \\
   &&  
     -8s^2 (s+t) J^{1}_{i8}  -2tu(5s+2t) J^{0}_{i12}-stu(3s+t)
J^{1}_{i17} 
      +tu (4s^2 +st-t^2 )J^{1}_{i21} 
          \nonumber \\
    &&\biggl. \Bigl. 
     -t^2 u J^{0}_{i21} \Bigr]     
  -F_{1}^{*} st \Bigl[ 4s J^{0}_{i2} +4t J^{0}_{i3} -3su J^{0}_{i17}
        -3tu J^{0}_{i21}     
        \Bigr] \biggr\}, \\
\Re( {\cal M}_{\Delta_-} {\cal M}_{O_c}^{\dagger} )&=&
        -\frac{g_s^6}{128\pi^4} C_1^2   
         \left(\frac{f_{\phi}}{4}\right)^2 8 m_b^2   
         \Re \biggl\{F_{3}^{*} \Bigl[
       s^2 (t-s-u) J^{0}_{i2}+st(t+u-s) J^{0}_{i3}
            \Bigr. \biggr. 
      \nonumber \\
   &&  \Bigl.
     -tu \Bigl( 2(s+t)J^{0}_{i12}  -s(2 J^{0}_{i8}+t J^{0}_{i21} 
      + (s +u)J^{0}_{i17} ) \Bigr) \Bigr] 
          \nonumber \\
    &&\biggl.  
    -F_{4}^{*} s^2 \Bigl[ 2(s+t) J^{0}_{i8} +tu(J^{0}_{i17}- J^{0}_{i21}
) \Bigr]     
  +F_{5}^{*} s \Bigl( s J^{0}_{i2} + t J^{0}_{i3}  \Bigr)     
       \biggr\}.
\end{eqnarray}
The functions $F_i$ are defined by 
\begin{eqnarray}
F_{0}&=&\int^1_0 dx \frac{\phi(x)m_b^4}{k_0^2 k^2},\\
 F_{1}&=&\int^1_0 dx \frac{x\phi(x)m_b^4}{k_0^2 k^2},\\
F_{3}&=&\int^1_0 dx \frac{\phi(x)m_b^2}{ k^2}G(r_3 ),\\
F_{4}&=&\int^1_0 dx \frac{x\phi(x)m_b^2}{ k^2}G(r_3 ),\\
F_{5}&=&\int^1_0 dx \frac{\phi(x)m_b^2}{ k_0^2}G(r_3 ),
\end{eqnarray}   
and $r_3 =\frac{k_0^2}{m_c^2}$. $\phi(x)$ is the distribution function
of $\phi$ 
meson, which can be found in Ref.~\cite{ball}. In numerical calculation
we take
$\phi(x)=6 x (1-x)$, which kills the IR pole of the gluon propagator
$\frac{1}{k^2}$. 
The other gluon is always time-like $k_0^2 > 0$, so there is no IR
divergence in our 
calculations.  

The $J$ functions are 
\begin{eqnarray}
J^{0}_{i2}=\int^1_0 dx r_0 \frac{\phi(x)\Delta i_2}{r_2}, &~~~&
 J^{0}_{i3}=\int^1_0 dx r_0 \frac{\phi(x)\Delta i_3}{r_2},\\
J^{0}_{i5}=\int^1_0 dx r_0 \frac{\phi(x)\Delta i_{5}}{r_2},&~~~& 
 J^{1}_{i5}=\int^1_0 dx r_0 \frac{x\phi(x)\Delta i_{5}}{r_2}, \\
J^{0}_{i6}=\int^1_0 dx r_0 \frac{\phi(x)\Delta i_{6}}{r_2},&~~~& 
J^{0}_{i8}=\int^1_0 dx r_0 \frac{\phi(x)\Delta i_{8}}{r_2},\\
 J^{1}_{i8}=\int^1_0 dx r_0 \frac{x\phi(x)\Delta i_{8}}{r_2}, &~~~&
J^{0}_{i11}=\int^1_0 dx r_0 \frac{\phi(x)\Delta i_{11}}{r_2},\\ 
 J^{0}_{i12}=\int^1_0 dx r_0 \frac{\phi(x)\Delta i_{12}}{r_2}, &~~~&
J^{0}_{i17}=\int^1_0 dx r_0^2 \frac{2\phi(x)\Delta i_{17}}{r_1 r_2},\\ 
 J^{1}_{i17}=\int^1_0 dx r_0^2 \frac{2x\phi(x)\Delta i_{17}}{r_1 r_2},
&~~~&
J^{0}_{i21}=\int^1_0 dx r_0^2 \frac{2\phi(x)\Delta i_{21}}{r_1 r_2}, \\
 J^{1}_{i21}=\int^1_0 dx r_0^2 \frac{2x\phi(x)\Delta i_{21}}{r_1
r_2},&~~~&
J^{0}_{i23}=\int^1_0 dx r_0^2 \frac{2\phi(x)\Delta i_{23}}{r_1 r_2}, \\
J^{1}_{i23}=\int^1_0 dx r_0^2 \frac{2x\phi(x)\Delta i_{23}}{r_1 r_2},
&~~~&
J^{2}_{i23}=\int^1_0 dx r_0^2 \frac{2x^2 \phi(x)\Delta i_{23}}{r_1 r_2},
\\
J^{0}_{i26}=\int^1_0 dx r_0^2 \frac{2\phi(x)\Delta i_{26}}{r_1 r_2},
&~~~&
J^{1}_{i26}=\int^1_0 dx r_0^2 \frac{2x\phi(x)\Delta i_{26}}{r_1 r_2}.
\end{eqnarray}

\end{document}